\documentclass{article}
\usepackage{graphicx}

\newcommand{\psect}[1]{{{\S}#1}}
\newcommand{\code}{\sf}
\newcommand{\el}[1]{{\code $<$#1$>$}}
\newcommand{\file}[1]{{\code #1}}
\newcommand{\p}{\par}
\newcommand{\ebibitem}{}
\newcommand{\url}[1]{{\code #1}}
\newcommand{\notex}[1]{}
\newcommand{\tex}[1]{{#1}}
\renewcommand{\title}[1]{\begin{center}{\LARGE #1}\end{center}}
\renewcommand{\author}[1]{\begin{center}{#1}\end{center}}

\begin{document}
\begin{flushright}
{\em High Energy Physics Libraries Webzine}, issue 4, June 2001\tex{\\}
\url{http://library.cern.ch/HEPLW/4/papers/3/}
\end{flushright}

\vspace*{1cm}

\title{Exposing and harvesting metadata using the OAI metadata\tex{\\} harvesting protocol: A tutorial}
\author{Simeon Warner \tex{(T-8, Los Alamos National Lab., USA)}}

\begin{abstract}
In this article I outline the ideas behind the Open Archives Initiative metadata 
harvesting protocol (OAIMH), and attempt to clarify some common misconceptions. 
I then consider how the OAIMH protocol can be used to expose and harvest metadata.
Perl code examples are given as practical illustration.
\end{abstract}


\section{Introduction}

The Open Archives Initiative (OAI)~\cite{OAI} announced the 
OAI metadata harvesting protocol (OAIMH) v1.0~\cite{OAIMH} on 21 January 2001
after a period of pre-release testing.
It is intended that the protocol will not be changed until 12-18 months
have elapsed from the initial release. This period of stability is designed 
to allow time for thorough evaluation without the cost of multiple rewrites
for early implementers. 
\p
The OAIMH protocol was designed as a simple, low-barrier way to
achieve interoperability through metadata harvesting. It is still an
open question as to exactly how useful metadata sharing will be.
However, there is certainly considerable interest in OAI and
experience with early OAIMH implementations is encouraging.
\p
This tutorial is organized in four main sections. In 
section~\ref{sec:not}, I hope to clear up some common misconceptions
about what OAIMH is. In section~\ref{sec:concepts}, I review some
of the concepts and assumptions that underly the OAIMH protocol.
Then, in the remaining two sections, sections~\ref{sec:dp} and~\ref{sec:sp}, 
I consider implementation of the {\em data-provider} and {\em service-provider}
sides of the OAIMH protocol. Perl code examples are given to implement
bare-bones versions of these two interfaces.
\p
It is not my intention to offer a complete description of the OAIMH protocol
but instead to describe its use in very practical terms, and to highlight
common practice among implementers.
A copy of the OAIMH protocol specification \cite{OAIMH} should be at
hand while reading this tutorial. I will refer to sections within
the protocol specification as \psect{2.1} (for section 2.1).

\section{\label{sec:not}What OAIMH is not}

The most common misconception of OAIMH, as it currently stands,
is that it provides mechanisms to expose and harvest 
full-content (documents, images, \ldots). This is not true, OAIMH 
is a protocol for the exchange of {\em metadata} only. However, it
may be that a future OAI protocol will provide facilities for
the exchange of full-content.
\p
OAIMH is not about direct interoperability between archives. It is based on
a model which puts a very clean divide between data-providers (entities which
expose metadata) and service-providers (entities which harvest metadata,
presumably with the intention of providing some service).
\p
\begin{figure}
\begin{center}
\includegraphics[height=5cm]{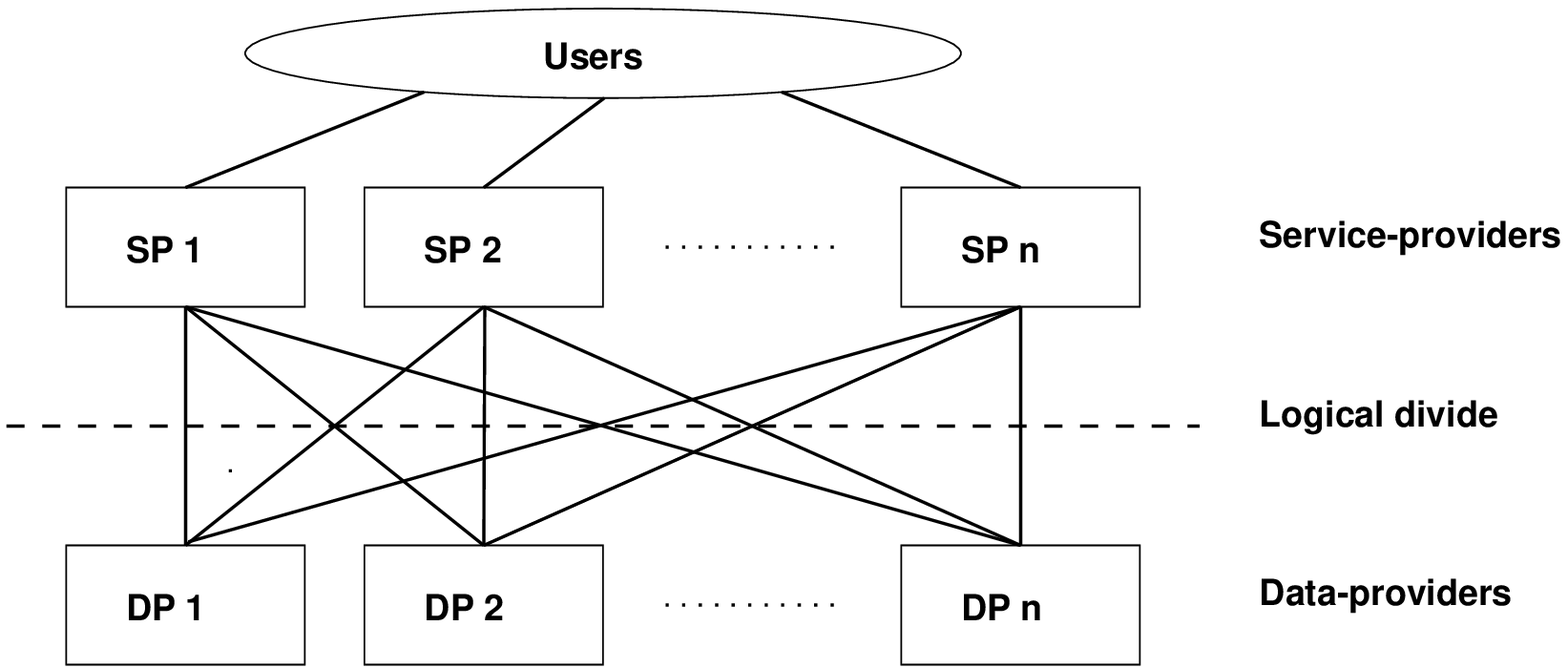}
\end{center}
\end{figure}
\p
While the model has a clear divide between data-providers and service-providers, 
there is nothing to say that one entity cannot be both; Cite Base~\cite{Citebase}
is one example. The model has an obvious scalability problem if every
service-provider is expected to harvest data from every data-provider. 
It may be that is is not an issue if service-providers are specific to
a particular community and thus harvest only from a subset of data-providers.
We may also see the creation of aggregators which harvest from a number of
data-providers and the re-export this data.
\p
OAIMH is not limited to Dublin Core (DC)~\cite{DC} metadata.
However, since OAI aims to promote interoperability, DC metadata 
has been adopted as a lowest common-denominator metadata format 
which all data-providers should support. 
It is not intended that the requirement to export DC metadata should
preclude the use of other metadata sets that may be more appropriate
within particular communities. The OAI encourages the development of
community specific standards that provide the functionalities required
by specific communities.


\section{\label{sec:concepts}OAIMH concepts}

\subsection{Pull-only interaction via HTTP using XML}

Service-providers make requests to data-providers; there is no support for
data-provider driven interaction. All requests and replies occur using the
HTTP protocol~\cite{HTTP}. Requests may be made using either the HTTP
GET or POST methods.
All successful replies are encoded in XML~\cite{XML}, and all exception and
flow-control replies are indicated by HTTP status codes.

\subsection{Verbs}

OAIMH protocol requests are made using one of six verbs: 
{\code Identify},
{\code GetRecord},
{\code ListIdentifiers},
{\code ListRecords},
{\code ListSets}, and
{\code ListMetadataFormats}.
Some of these verbs accept or require additional parameters to
completely specify the request. The verb and any parameters are
specified as {\code key=value} pairs~\psect{3.1.1} either in the URL
(if the GET method is used), or in the body of the message (if the
POST method is used). 

\subsection{Items, records and identifiers}

The OAIMH protocol is based on a model of repositories that hold metadata about 
{\em items}~\psect{2}. The nature of the items is outside the scope of the protocol; they
might be electronic documents, artifacts in a museum, people, or almost anything
else. The requirement for OAI compliance is that the repository be able
to disseminate metadata for these items in one or more formats including
Dublin Core (DC).
\p
Metadata is disseminated via the {\code GetRecord} and {\code ListRecords}
verbs. These requests result in zero or more {\em records} being returned.
A record consists of 2 or 3 parts: a \el{header} container, a \el{metadata}
container, and possibly an \el{about} container~\psect{2.2}.
\p
The metadata for each item has a unique identifier which, when combined
with the {\code metadataPrefix}, acts as a key to extract a metadata record.
Note that although all metadata types for an item share the same identifier,
the identifier is  explicitly {\em not} an identifier for the item~\psect{2.3}.
Identifiers may be any valid URI~\cite{URI} but an optional OAI identifier
syntax \psect{A2} has been adopted widely. 
The OAI identifier syntax divides the identifier into three parts
separated by colons (:), e.g. {\code oai:arXiv:hep-lat/0008015} where 
`{\code oai}' is the scheme, `{\code arXiv}' identifies the repository, 
and `{\code hep-lat/0008015}' is the identifier within the particular repository.   

\subsection{Datestamps}

The metadata for an item (considered as a whole, not as individual formats)
has a datestamp which is the date of last modification. The purpose of the datestamp
is to support date-selective harvesting and incremental harvesting in particular.
Datestamps are returned in the records returned by a data-provider and may be used
as optional arguments to the {\code ListIdentifiers} and {\code ListRecords}
requests.
\p
The datestamps have the granularity of a day, they are in YYYY-MM-DD format and
no time is specified. This simple date format actually creates some additional
complexity because the service-provider and data-provider may not be in the 
same time-zones. This is considered further in section~\ref{sec:incHarvest}.
\p
Typically, a service-provider would initially harvest all metadata records from a 
repository by issuing a {\code ListRecords} request without {\code from} or
{\code until} restrictions. Subsequently, the service-provider would issue
{\code ListRecords} requests with a {\code from} parameter equal to the date
of the last harvest.  

\subsection{Sets}

Sets are provided as an {\em optional} construct for grouping items to support
selective harvesting \psect{2.5}. It is not intended that they should provide
a mechanism by which a search is implemented, and there is no controlled vocabulary
for set names so automated interpretation of set structure is not supported.
It should be noted that sets are optional both from the point of view of
the data-provider --- which may or may not implement sets; and 
the service-provider --- which may ignore any set structure that is exposed.
It is not clear whether sets will be widely used and I shall not consider them
further in this tutorial.
 
\subsection{Metadata formats}

The OAIMH protocol supports {\em multiple parallel metadata formats}. Dublin Core (DC)
is mandated for lowest common denominator interoperability. The use of other formats
within particular communities or for special purposes is encouraged. Within a 
particular repository, metadata formats are identified by a {\code metadataPrefix}.
Each {\code metadataPrefix} is associated with the URL of the schema which may
be used to validate the metadata records; the URL has cross repository scope.  
The only globally meaningful {\code metadataPrefix} is {\code oai\_dc} (for DC),
which is associated with the schema at 
{\code http://www.openarchives.org/OAI/dc.xsd}.   
\p
The {\code ListMetadataFormats} request will return the {\code metadataPrefix}, 
{\code schema}, and optionally a {\code metadataNamespace}, for either 
a particular record or for the whole repository (if no identifier is specified).
In the case of the
whole repository, all metadata formats supported by the repository are returned. 
It is not implied that all records are available in all formats.   

\subsection{Exception conditions}

The OAIMH protocol has very simple exception handling: syntax errors result in 
HTTP status code 400 replies, and parameters that are invalid or have values that do
not match records in the repository result in empty replies. For example, a
{\code ListRecords} request for a date range when there were no changes, or
for a metadata format not supported, will result in a reply with header information
but no \el{record} elements.

\subsection{Flow control, load balancing and redirection}

Flow control is supported with the HTTP retry-after status code 503. This allows
a server (data-provider) to tell the harvesting agent (service-provider) to 
try the request again after some interval. It is left entirely up to the 
server implementer to determine the conditions under which such a response 
will be given. The server could base the response on current machine load or
limit the frequency requests will be serviced from any given IP address. 
The retry-after response may also be used to handle temporary outages without
simple taking the server off-line.
\p
In an environment where one of a set of servers may handle a request, the
server may dynamically redirect a request using the HTTP 302 response. To date
this has been implemented only by the NACA repository~\cite{NACA}.


\section{\label{sec:dp}Exposing metadata}

To expose metadata within the OAI, one must implement the
data-provider site of the OAIMH protocol. This provides a small set
of functions which can be used to extract information about and metadata 
from the underlying repository.

\subsection{Minimal server implementation}

The Perl files \file{oai1.pl}, \file{OAIServer.pm} and \file{Database.pm} 
implement a bare-bones data-provider interface. The file \file{oai1.pl}
handles HTTP requests and must be associated with a URL in the
web server configuration file; for the Apache~\cite{Apache} web server, 
the configuration line is {\code ScriptAlias /oai1 /some/directory/oai1.pl}
if the code is in {\code /some/directory}. It is also possible to run \file{oai1.pl}
from the command line, the request is specified with the {\code -r}
flag, e.g. {\code ./oai1.pl -r 'verb=Identify'}.
\p
The algorithm for \file{oai1.pl} is simply:
\begin{verbatim}
read GET, POST or command line request
check syntax of request
if syntax correct 
  return XML reply to request
else
  return HTTP 400 error code and message 
\end{verbatim}
\p
An example of an invalid request is:
\begin{verbatim}
simeon@fff>./oai1.pl -r 'bad-request'
Status: 400 Malformed request
Content-Type: text/plain

No verb specified!
\end{verbatim}
\p
\file{OAIServer.pm} exports two subroutines, one ({\code OAICheckRequest}) 
to check the request against a grammar stored in a data structure, and 
another ({\code OAISatisfyRequest}) which calls the appropriate routine
to implement the required OAI verb. I will consider each verb in turn.
\p
\file{Database.pm} is a dummy database interface
with a `database' of three records: {\code record1}, {\code record2} and
{\code record3}. Metadata for {\code record1} and {\code record2} is available
in DC format; metadata for {\code record1} is also available in another
format with the {\code metadataPrefix} `{\code wibble}'; and {\code record3} is a
`deleted' record so no metadata is available.

\subsection{Identify}

This verb takes no arguments and returns information about a 
repository~\psect{4.2}. The example code implements Identify by
simply writing out information from configuration variables. The
protocol allows for additional \el{description} blocks which may
contain community-specific information. Examples include
\el{oai-identifier} which specifies a particular identifier syntax,
and \el{eprints} which includes additional information appropriate
for the e-print community~\psect{A2}.

\subsection{ListSets}

This verb takes no arguments and returns the set structure of the 
repository~\psect{4.6}. The example code does not implement any
sets so the response is an empty list.
  
\subsection{ListMetadataFormats}

This verb may be used either with a {\code identifier} argument or without
any arguments~\psect{4.4}. If an {\code identifier} is specified then 
the verb returns the metadata formats available for that record. In many
cases a repository may be able to disseminate metadata for all records in the
same format or formats. In this case the response will be the same if
there is no {\code identifier} argument or if the {\code identifier} argument
specifies any record that exists.
The example code implements the general case by calling a routine in the
\file{Database.pm} module to ask what formats are available, and then
formats the reply appropriately. For each metadata format, the reply
must include a \el{metadataPrefix} (used to specify that format in other 
requests), and a \el{schema} URL. A \el{metadataNamespace} element may
optionally be returned but is not implemented in the example code.
\p
An example request and response is: 
\begin{verbatim}
simeon@fff>./oai1.pl -r 'verb=ListMetadataFormats&identifier=record1'
Content-Type: text/xml

<?xml version="1.0" encoding="UTF-8"?>

<ListMetadataFormats xmlns="http://www.openarchives.org/OAI/OAI_ListMetadataFormats"
    xsi:schemaLocation="http://www.openarchives.org/OAI/1.0/OAI_ListMetadataFormats
                        http://www.openarchives.org/OAI/1.0/OAI_ListMetadataFormats.xsd">
 <responseDate>2001-05-05T12:27:36-06:00</responseDate>
 <requestURL>http://localhost/oai1?verb=ListMetadataFormats&amp;
             identifier=record1&amp;verb=ListMetadataFormats</requestURL>
 <metadataFormat>
  <metadataPrefix>wibble</metadataPrefix>
  <schema>http://wibble.org/wibble.xsd</schema>
 </metadataFormat>
 <metadataFormat>
  <metadataPrefix>oai_dc</metadataPrefix>
  <schema>http://www.openarchives.org/OAI/dc.xsd</schema>
 </metadataFormat>
</ListMetadataFormats>
\end{verbatim}
The response indicates that the record {\code record1} may
be disseminated in either {\code oai\_dc} or {\code wibble} formats.

\subsection{GetRecord}

This verb requests metadata for a particular record in a particular
format~\psect{4.1}. The example code implements this as a call to
a subroutine {\code disseminate} (shared with {\code ListRecords})
after checking that the record exists.
\p
The record returned consists of two parts if the record is not
deleted; a \el{header} block which contains the identifier and the datestamp
(the information required for harvesting) and a \el{metadata} block
which contains the XML metadata record in the requested format. The 
\el{metadata} block will be missing if the record is deleted or if the
requested metadata format is not available.
\p
For example, a request for {\code oai\_dc} for {\code record2} would
be:
\begin{verbatim}
simeon@fff>./oai1.pl -r 'verb=GetRecord&identifier=record2&metadataPrefix=oai_dc'
Content-Type: text/xml

<?xml version="1.0" encoding="UTF-8"?>

<GetRecord xmlns="http://www.openarchives.org/OAI/OAI_GetRecord"
  xsi:schemaLocation="http://www.openarchives.org/OAI/1.0/OAI_GetRecord
                      http://www.openarchives.org/OAI/1.0/OAI_GetRecord.xsd">
 <responseDate>2001-05-05T12:50:23-06:00</responseDate>
 <requestURL>http://localhost/oai1?verb=GetRecord&amp;identifier=record2&amp;
             metadataPrefix=oai_dc&amp;verb=GetRecord</requestURL>
 <record>
  <header>
   <identifier>record2</identifier>
   <datestamp>1999-02-12</datestamp>
  </header>
  <metadata>
   <oai_dc xsi:schemaLocation="http://purl.org/dc/elements/1.1/
                               http://www.openarchives.org/OAI/dc.xsd"
	   xmlns:xsi="http://www.w3.org/2000/10/XMLSchema-instance"
	   xmlns="http://purl.org/dc/elements/1.1/">
    <title>Item 2</title>
    <creator>A N Other</creator>
   </oai_dc>
  </metadata>
 </record>
</GetRecord>
\end{verbatim}
but a request for the unavailable format {\code wibble} would be:
\begin{verbatim}
simeon@fff>./oai1.pl -r 'verb=GetRecord&identifier=record2&metadataPrefix=wibble'
Content-Type: text/xml

<?xml version="1.0" encoding="UTF-8"?>

<GetRecord xmlns="http://www.openarchives.org/OAI/OAI_GetRecord"
 xsi:schemaLocation="http://www.openarchives.org/OAI/1.0/OAI_GetRecord
                     http://www.openarchives.org/OAI/1.0/OAI_GetRecord.xsd">
 <responseDate>2001-05-05T12:52:13-06:00</responseDate>
 <requestURL>http://localhost/oai1?verb=GetRecord&amp;
   identifier=record2&amp;metadataPrefix=wibble&amp;verb=GetRecord</requestURL>
 <record>
  <header>
   <identifier>record2</identifier>
   <datestamp>1999-02-12</datestamp>
  </header>
 </record>
</GetRecord>
\end{verbatim}
which includes a \el{header} block but no \el{metadata} block.
\p
The protocol also permits the addition of an \el{about} container~\psect{2.2} 
for each record This is provided as a hook for additional
information such as rights or terms information. It is not currently used
by any of the registered OAI data-providers and is not implemented in the
example code.

\subsection{ListIdentifiers and ListRecords}

ListIdentifiers~\psect{4.3} and ListRecords~\psect{4.5} both implement
a search by date, the difference is whether they return a list of
identifiers or complete metadata records in the specified format. 
The example code implements both of these verbs using the subroutine
{\code listEither} which calls a search by date ({\code getIdsByDate})
in \file{Database.pm}.
\p
In the case of ListIdentifiers the output consists of \el{identifier}
elements which may include the attribute {\code status="deleted"} if
the record is deleted. An example request without date restriction
is:
\begin{verbatim}
simeon@fff>./oai1.pl -r 'verb=ListIdentifiers'
Content-Type: text/xml

<?xml version="1.0" encoding="UTF-8"?>

<ListIdentifiers xmlns="http://www.openarchives.org/OAI/OAI_ListIdentifiers"
  xsi:schemaLocation="http://www.openarchives.org/OAI/1.0/OAI_ListIdentifiers
                      http://www.openarchives.org/OAI/1.0/OAI_ListIdentifiers.xsd">
 <responseDate>2001-05-05T12:59:30-06:00</responseDate>
 <requestURL>http://localhost/oai1?verb=ListIdentifiers&amp;verb=ListIdentifiers</requestURL>
 <identifier>record1</identifier>
 <identifier>record2</identifier>
 <identifier status="deleted">record3</identifier>
</ListIdentifiers>
\end{verbatim}
The response lists the identifiers of the three records in the repository and
indicates that {\code record3} is deleted.
If the parameter {\code until=2000-01-01} were added then only the
first two identifiers would be returned since the datestamp of {\code record3}
is 2000-03-13.
\p
In the case of {\code ListRecords} the output consists of \el{record} blocks
similar to those obtained from GetRecord requests. {\code ListRecords} requests
must include a {\code metadataPrefix} parameter.

\subsection{Partial responses}

The OAIMH protocol allows for partial responses~\psect{3.4} for all of the list
verbs ({\code ListIdentifiers}, {\code ListSets}, {\code ListMetadataFormats}
and {\code ListRecords}). This feature has been implemented by most of the larger
registered OAI repositories for the {\code ListIdentifiers} and {\code ListRecords}
verbs. The example code does not implement this feature.


\section{\label{sec:sp}Harvesting metadata}

To harvest metadata within the OAI, one must implement the
service-provider site of the OAIMH protocol.
I will consider the implementation of a harvester that performs two
functions: firstly, harvest all metadata in a particular format, and
secondly, harvest all metadata in a particular format that has changed
since a given date. These functions are the basis of a system that can
create and maintain an up to date copy of the metadata from an OAI
compliant repository.
\p
As one of the maintainers of a heavily used archive I am painfully aware 
of the importance of avoiding inadvertent denial-of-service attacks created
by badly written harvesting software. Automated agents should always include
a useful user-agent string and a valid e-mail contact address in their
HTTP requests. The flow-control elements of the protocol must be respected
and careful testing is essential.

\subsection{Detecting changes that require manual intervention}

I will assume that the goal is to create software which will
run on some schedule so that the local copy of metadata from some
set of repositories is kept updated without manual intervention.
However, it would be reckless to assume that the details of 
repositories will not change over time. In order to avoid the need
for manual polling to detect such changes, we should ask how they
can be detected automatically.
\p
To detect changes other than the addition and deletion of records 
which are part of normal repository operation, one can compare the
response to OAI requests that describe the repository between
successive harvests. These requests are {\code Identify} and
probably {\code ListSets} and {\code ListMetadataFormats} (for the
whole repository as opposed to any single record). For all of the
requests we expect the \el{responseDate} to change with each request
but for these requests we expect the rest of the response to be 
unchanged. Note that to do the test correctly one should compare the
XML data in such a way that valid transformations, say re-ordering
elements, are ignored. However, in practice it is likely to be 
sufficient (if over sensitive) to do a string comparison
of the responses so long as changes in the \el{responseDate} are ignored.
\p
In the example harvester I have included the facility to specify
a file containing the {\code Identify} response from the previous harvest.
This is used both to extract the date of the last harvest and
to check for changes in that response. I have not implemented a
test for changes in the {\code ListSets} and
{\code ListMetadataFormats} responses.

\subsection{\label{sec:incHarvest}Incremental harvesting}

The OAIMH protocol was designed to facilitate incremental harvesting.
The idea is that a service-provider will maintain an up to date copy
of the metadata from a repository by periodically harvesting 
`changed' records. This is why all records have a datestamp, the date
of last modification, associated with them.
\p
The 1 day granularity of the datestamp and the possibility of
data-providers and service-providers being in different time zones means
that there must be some overlap between the date ranges of successive 
requests~\cite{harvestingStrategy}.
If the service-provider and data-provider share the same time-zone then
a 1~day overlap is sufficient to ensure that updates are not missed; 
records might be updated after the harvest on the day of the last harvest,
but provided records that have changed on that day are reharvested then no changes will
be missed. To cope with different time zones it is necessary to extend
this to a 2~day overlap if the harvester works with dates local to
itself. An alternative strategy, which I prefer, is to use only the
dates returned by the repository and thus, by working in the local
time zone of the repository, reduce the required overlap to 1~day. 
\p
In the example harvester I implement this last strategy by taking the
date of the last harvest from the \el{responseDate} of the stored
{\code Identity} response (the \el{responseDate} must be specified in the local 
time zone of the repository~\psect{3.1.2.1}. This date may then be
used as the  {\code from} date (inclusive) for the next {\code ListRecords}
or {\code ListIdentifiers} request.

\subsection{Flow control and redirection}

The module \file{OAIGet.pm} examines the HTTP reply for status codes
302 (redirect) and 503 (retry-after). Both replies are handled 
automatically, a default retry period is assumed if the 503 response
does not specify a time (though this is an error on the part of the
data-provider). Messages are printed if the verbose option is selected.

\subsection{Parsing replies}

OAIMH protocol replies are designed to be self contained, in part to
allow off-line processing thereby separating the harvesting and 
database-update processes.
However, in order to deal with partial responses the harvesting software
must be able to parse the responses to all the list requests
({\code ListIdentifiers}, {\code ListSets}, {\code ListMetadataFormats} and
{\code ListRecords}) sufficiently to extract any {\code resumptionToken}~\psect{3.4}.
To date, none of the registered OAI compliant repositories give
partial responses for {\code ListSets} and {\code ListMetadataFormats} 
requests, but several do for {\code ListIdentifiers} and 
{\code ListRecords} requests. 
\p
Perhaps the neatest way to implement a harvester would be to have it
recombine partial responses into a complete reply. The example code
does not do this but does parse all list requests to look for a 
\el{resumptionToken} so that further requests can be used to complete the
original request. 

\subsection{An example harvester}

The files \file{oaiharvest.pl}, \file{OAIGet.pm} and \file{OAIParser.pm}
implement a simple harvester that illustrates the points mentioned
above. \file{oaiharvest.pl} is the executable and accepts a variety
of flags, these can be displayed by executing {\code oaiharvest.pl -h}.
The algorithm is:
\begin{verbatim}
read command line arguments
check options and parameters
issue Identify request
compare response with previous Identify response if given
extract `from' date from command line, previous Identify response or do complete harvest
LOOP:
  issue ListRecords or ListIdentifiers request
  check for resumptionToken, LOOP if present
\end{verbatim}
\p
The subroutine {\code OAIGet} in \file{OAIGet.pm} is used to issue the 
OAIMH requests and this handles any retry-after or redirect replies.
XML parsing is handled by the \file{OAIParser.pm} module which extends
{\code XML-Parser}, which itself is based on the {\code expat} parser.
\p
Let us take as an example, harvesting metadata from the example data-provider
code which has be set up at the URL {\code http://localhost/oai1}. First
we would issue a harvest command without any time restriction (to harvest
all records). In the examples, I harvest just the identifiers using
{\code ListIdentifiers} requests, the flags {\code -r} and {\code -m metadataPrefix}
can be used to instruct \file{oaiharvest.pl} to issue {\code ListRecords} 
requests and to specify a {\code metadataPrefix} other than {\code oai\_dc}.
\p
\begin{verbatim}
simeon@fff>mkdir harvest1
simeon@fff>./oaiharvest.pl -d harvest1 http://localhost/oai1

oaiharvest.pl: Harvest from http://localhost/oai1 using POST
OAIGet: Doing POST to http://localhost/oai1 args: verb=Identify
OAIGet: Got 200 OK (479bytes)
oaiharvest.pl: Doing complete harvest.
OAIGet: Doing POST to http://localhost/oai1 args: verb=ListIdentifiers
OAIGet: Got 200 OK (537bytes)
oaiharvest.pl: Got 3 identifiers (running total: 3)
oaiharvest.pl: No resumptionToken, request complete.
oaiharvest.pl: Done.

simeon@fff>ls harvest1
Identify  ListIdentifiers.1
\end{verbatim}
\p
If we then do an incremental harvest specifying the file name of
the last {\code Identify} response, {\code harvest1/Identify}, the harvester
checks against this response for changes (none except date) and extracts
the date of the last harvest (2001-06-05) to be used as the {\code from}
date for the new harvest. 
\begin{verbatim}
simeon@fff>mkdir harvest2
simeon@fff>./oaiharvest.pl -d harvest2 -i harvest1/Identify  http://localhost/oai1

oaiharvest.pl: Harvest from http://localhost/oai1 using POST
OAIGet: Doing POST to http://localhost/oai1 args: verb=Identify
OAIGet: Got 200 OK (479bytes)
oaiharvest.pl: Identify response unchanged from reference (except date)
oaiharvest.pl: Reading harvest1/Identify to get from date
oaiharvest.pl: Incremental harvest from 2001-06-05 (from harvest1/Identify)
OAIGet: Doing POST to http://localhost/oai1 args: from=2001-06-05&verb=ListIdentifiers
OAIGet: Got 200 OK (444bytes)
oaiharvest.pl: Got 0 identifiers (running total: 0)
oaiharvest.pl: No resumptionToken, request complete.
oaiharvest.pl: Done.
\end{verbatim}
Since there have been no changes in the database this harvest results in no
identifiers being returned.
\p
To extend this example, I then edited the database (\file{Database.pm}) to
add a new record ({\code record4}) with datestamp {\code 2001-06-05} which
simulates the addition of a record after the last harvest but on the same day.
I then ran another harvest command.
\begin{verbatim}
simeon@fff>diff Database.pm~ Database.pm
24c24,26
<   'record3' => [ '2000-03-13', undef ]  #deleted
---
>   'record3' => [ '2000-03-13', undef ],  #deleted
>   'record4' => [ '2001-06-05', {
>     'oai_dc' => ['title','Item 4', 'creator','Someone Else'] } ] 

simeon@fff>mkdir harvest3
simeon@fff>./oaiharvest.pl -d harvest3 -i harvest2/Identify http://localhost/oai1

oaiharvest.pl: Harvest from http://localhost/oai1 using POST
OAIGet: Doing POST to http://localhost/oai1 args: verb=Identify
OAIGet: Got 200 OK (479bytes)
oaiharvest.pl: Identify response unchanged from reference (except date)
oaiharvest.pl: Reading harvest2/Identify to get from date
oaiharvest.pl: Incremental harvest from 2001-06-05 (from harvest2/Identify)
OAIGet: Doing POST to http://localhost/oai1 args: from=2001-06-05&verb=ListIdentifiers
OAIGet: Got 200 OK (478bytes)
oaiharvest.pl: Got 1 identifiers (running total: 1)
oaiharvest.pl: No resumptionToken, request complete.
oaiharvest.pl: Done.
\end{verbatim}
This harvest results in one additional identifier, {\code record4}, being returned as expected.
\p
Below are two excerpts from harvests from real repositories which 
illustrate the flow-control features of the protocol. The first is from
{\code arXiv} which uses 503 retry-after replies to enforce a delay
between requests. The second if from {\code NACA} which uses 302 redirect
replies to demonstrate a load-sharing scheme.
\begin{verbatim}
...
OAIGet: Doing POST to http://arXiv.org/oai1 args: verb=ListIdentifiers
OAIGet: Got 503, sleeping for 60 seconds...
OAIGet: Woken again, retrying...
OAIGet: Got 200 OK (27398bytes)
oaiharvest.pl: Got 502 identifiers (running total: 502)
oaiharvest.pl: Got resumptionToken: `1997-02-10___'
OAIGet: Doing POST to http://arXiv.org/oai1 args: resumptionToken=1997-02-10___&verb=ListIdentifiers
OAIGet: Got 503, sleeping for 60 seconds...
OAIGet: Woken again, retrying...
OAIGet: Got 200 OK (28330bytes)
oaiharvest.pl: Got 520 identifiers (running total: 1022)
oaiharvest.pl: Got resumptionToken: `1997-03-06___'
...

...
OAIGet: Doing POST to http://naca.larc.nasa.gov/oai/ args: verb=ListIdentifiers
OAIGet: Got 302, redirecting to http://buckets.dsi.internet2.edu/naca/oai/?...
OAIGet: Doing POST to http://buckets.dsi.internet2.edu/naca/oai/ args: verb=ListIdentifiers
OAIGet: Got 200 OK (336705bytes)
oaiharvest.pl: Got 6352 identifiers (running total: 6352)
...
\end{verbatim}
\p
I hope the examples above provide a useful demonstration of some of the features of
the OAIMH metadata harvesting. Be sure to exercise caution and restraint when
running tests against registered repositories. There is some cost in associated
with answering OAIMH requests, and recklessly downloading large amounts of data for
no good reason is not helpful.


\section{Conclusions} 

The OAIMH protocol has been public for 5 months now and experience shows that
it is adequate for its intended purpose. There are now 30 registered repositories
which together expose over 600,000 metadata records. While there are currently
just two registered service providers, `arc'~\cite{arc} and the Repository 
Explorer~\cite{repositoryExplorer}, there is an increasing number of
tools and libraries available to assist in the development of harvesting
applications. Publicly available tools and libraries are listed on the
OAI web site~\cite{OAIsoftware}. This includes Tim Brody's~\cite{TimBrodyLibrary}
Perl library which is considerably more extensive than the examples presented here.
\p
The uptake of OAI is very encouraging and it is feedback from the current
implementers which will shape the next version of the OAIMH protocol.
Anyone implementing, or interested in implementing, either side of the OAIMH
protocol should subscribe to the {\code oai-implementers}~\cite{oaiImplementers}
mailing list. It is a helpful and friendly forum.
 

\section*{Appendix: Example programs}

The example programs are:
\begin{itemize}
\item \file{oai1.pl} and \file{OAIServer.pm} for the server; and
\item \file{oaiharvest.pl}, \file{OAIGet.pm} and \file{OAIParser.pm} for the harvester.
\end{itemize}
\notex{These files are available as a gzipped tar archive: \file{examples.tar.gz} (10kB), or as a zip archive: \file{examples.zip} (12kB).}
\tex{These files are included with this paper, please download the source.}
\p
In order to run the example programs, you will require Perl 5.004 
or later and the following modules
(the precise version I used is given in parenthesis). For the
the server:
\begin{itemize}
\item XML-Writer (XML-Writer-0.4)
\end{itemize}
and for the harvester:
\begin{itemize}
\item MIME-Base64 (MIME-Base64-2.11)
\item URI (URI-1.09)
\item HTML-Tagset (HTML-Tagset-3.02)
\item HTML-Parser (HTML-Parser-3.11)
\item libnet (libnet-1.0703)
\item Digest::MD5 (Digest-MD5-2.11)
\item LWP (libwww-perl-5.48)
\item expat library (expat-1.95.1)
\item XML-Parser (XML-Parser-2.30)
\end{itemize}
All of the above except for {\code expat} are available from CPAN 
({\code http://www.cpan.org/}) and can be installed with the standard 
{\code perl Makefile.PL; make; make test; make install} sequence.
There should not be any dependency problems if the modules are installed 
in the order listed.
The {\code expat} XML parsing library upon which {\code XML-Parser} relies,
is available from Source Forge ({\code http://sourceforge.net/projects/expat/}).
\p 
Before running \file{oaiharvest.pl} you should first edit the line that
defines the variable {\code \$contact} and insert your e-mail address. This
will then be specified as the contact address for all HTTP requests and
will enable the server maintainer to contact you if there are problems.
The example code has been tested only on a Linux system and with the
Apache server. While I hope that it will work on other systems 
this has not been verified.


\section*{About the author} 

Simeon Warner is one of the maintainers and developers of the 
arXiv e-print archive. He has been actively involved with the
development and implementation of the OAI since its inception.
\p
address: T-8, Los Alamos National Lab., Los Alamos, NM 87545, USA\\
e-mail: simeon@lanl.gov



\begin{thebibliography}{1}

\bibitem{OAI}
Open Archives Initiative (OAI),
\newblock \url{http://www.openarchives.org/}
\ebibitem

\bibitem{OAIMH}
OAI metadata harvesting protocol v1.0, released on 21~January~2001, revised 24~April~2001
\newblock \url{http://www.openarchives.org/OAI/openarchivesprotocol1.0.html}
\ebibitem

\bibitem{Citebase}
Cite Base at the University of Southampton,
\newblock a prototype Open Archives federating service which extracts and re-exports citation information in addition to providing a search facility,
\newblock \url{http://cite-base.ecs.soton.ac.uk/}
\ebibitem

\bibitem{DC}
Dublin Core Metadata Element Set, Version~1.1: Reference Description (2~July~1999),
\newblock \url{http://purl.org/DC/documents/rec-dces-19990702.htm}

\bibitem{HTTP}
HTTP - Hypertext Transfer Protocol v1.1, 
\newblock \url{http://www.ietf.org/rfc/rfc2616.txt}
\ebibitem

\bibitem{XML}
XML - Extensible Markup Language,
\newblock OAI uses XML schemas to specify responses,
\newblock \url{http://www.w3.org/XML/}
\newblock \url{http://www.w3.org/TR/xmlschema-0/} 
\ebibitem

\bibitem{URI}
\newblock Uniform Resource Identifiers (URI): Generic Syntax,
\newblock \url{http://www.ietf.org/rfc/rfc2396.txt}
\ebibitem

\bibitem{NACA}
NACA - National Advisory Committee for Aeronautics Technical Report Server,
\newblock \url{http://naca.larc.nasa.gov/}
\ebibitem

\bibitem{Apache}
The Apache web server
\newblock \url{http://www.apache.org/}
\ebibitem

\bibitem{harvestingStrategy}
Harvesting strategies have been discussed on the {\code oai-implementers} list~\cite{oaiImplementers},
\newblock I have drawn from the comments of Hussein Suleman in particular.
\ebibitem

\bibitem{arc}
`arc' Cross Archive Searching Service,
\newblock an OAI {\em service provider} developed at Old Dominion University,
\newblock \url{http://arc.cs.odu.edu/help/archives.htm}
\ebibitem

\bibitem{repositoryExplorer}
The OAI Repository Explorer,
\newblock an interface to interactively test archives for compliance with the OAIMH protocol,
\newblock Hussein Suleman (Digital Libraries Research Laboratory, Virginia Tech.),
\newblock \url{http://rocky.dlib.vt.edu/~oai/cgi-bin/Explorer/oai1.0/testoai}
\ebibitem

\bibitem{OAIsoftware}
Open Archives Initiative, Tools for Implementers list, 
\newblock \url{http://www.openarchives.org/tools/tools.htm}
\ebibitem

\bibitem{TimBrodyLibrary}
Perl class library that allow the rapid deployment of an OAI compatible interface to an existing web server/database for OAI server and harvester implementation,
\newblock \url{http://www.ecs.soton.ac.uk/\~tdb198/oai/frontend.html}
\ebibitem

\bibitem{oaiImplementers}
{\code oai-implementers},
\newblock a mailing list (and archive) for discussing the implementation of the OAIMH protocol,
\newblock \url{http://oaisrv.nsdl.cornell.edu/mailman/listinfo/OAI-implementers}
\ebibitem

\end{thebibliography}
\end{document}